\documentclass[conference,compsoc]{IEEEtran}

\usepackage[nocompress]{cite}
\usepackage[utf8]{inputenc}
\usepackage{tikz}
\usetikzlibrary{positioning}
\usetikzlibrary{calc}
\usetikzlibrary{chains}
\usetikzlibrary{shapes}
\usepackage{natbib}
\usepackage{booktabs}
\usepackage{graphics}
\usepackage{wrapfig}
\usepackage{placeins}
\usepackage{amsmath}
\usepackage{listings}
\usepackage{microtype}
\usepackage{hyperref}
\usepackage{hypernat}
\usepackage{ifthen}
\usepackage{amssymb}
\usepackage{algorithm}
\usepackage{algorithmic}

\newboolean{doubleblind}
\setboolean{doubleblind}{False}

\def\loopy{Loopy}
\def\gsize{\text{gsize}}

\newlength\q
\definecolor{green}{RGB}{0, 180, 0}

\lstdefinestyle{custompython}{%
    belowcaptionskip=1\baselineskip,
    breaklines=true,
    frame=none,
    xleftmargin=\parindent,
    language=Python,
    showstringspaces=false,
    basicstyle=\footnotesize\ttfamily,
    keywordstyle=\bfseries\color{green!40!black},
    commentstyle=\itshape\color{purple},
    identifierstyle=\color{black},
    stringstyle=\color{green!60!black},
    keywords=[2]{as,True,False},
    keywordstyle=[2]\bfseries\color{green!40!black},
    keywords=[3]{np,sp,plt},
    keywordstyle=[3]\bfseries\color{blue},
    numbers=none,
    columns=fullflexible,
}

\title{A Unified, Hardware-Fitted, Cross-GPU Performance Model}

\ifthenelse{\boolean{doubleblind}}{
  \author{*} 
}{
  \author{James D. Stevens\\
  Andreas Klöckner\\[.5cm]
  Department of Computer Science\\
  University of Illinois at Urbana-Champaign}
}

\begin{document}


\maketitle

\begin{abstract}
  We present a mechanism to symbolically gather performance-relevant operation
  counts from numerically-oriented subprograms (`kernels') expressed in the
  \loopy\ programming system, and apply these counts in a simple, linear model
  of kernel run time. We use a series of `performance-instructive' kernels to
  fit the parameters of a unified model to the performance characteristics of
  GPU hardware from multiple hardware generations and vendors. We evaluate the
  predictive power of the model on a broad array of computational kernels
  relevant to scientific computing. In terms of the geometric mean, our simple,
  vendor- and GPU-type-independent model achieves relative accuracy comparable
  to that of previously published work using hardware specific models.

\end{abstract}


\section{Introduction}
\label{sec:intro}


Being able to approximately predict the running time of computational kernels is
a key step towards the automation of performance tuning for complicated, modern,
vector-based, massively parallel processor architectures.  We present a simple,
effective model to achieve such a prediction that is realized on top of, though
technically not dependent on, a
program transformation system, providing a self-contained foundational building
block to aid the developer of automated tuning solutions in exploring the vast
search space of possible and, from the point of view of the result, equivalent program
variants. We note that we mainly view our model as a more economical alternative
to evaluating the execution time of a computational kernel than, for example,
using actual on-device timing runs. Our system primarily targets the execution
paradigm embodied by modern GPU Hardware, as exposed in, for example, the CUDA
or OpenCL compute abstractions. The system makes no assumptions about the
internal organization of the hardware, and device-specific parameters are
obtained from a black-box adaptation process that needs to run precisely once on
each new piece of hardware on which the system is used.

GPUs, originally designed for rapid graphics rendering, have highly parallel
single instruction, multiple data (SIMD) architectures that make them
particularly useful for data-parallel problems. Over the last decade, general
purpose GPU programming has risen in popularity. Some of the world's fastest
supercomputers~\citep{meuer2015top500}, make use of thousands of GPU nodes, including Oak Ridge National
Laboratory's Titan supercomputer. GPU programming has been facilitated by the
release of general purpose GPU programming systems, including Nvidia CUDA in 2007
and the Open Computing Language (OpenCL)
in 2009 \citep{nvidia2015cuda,munshi2011opencl}.

Much of the previous work in GPU performance modeling has focused on
constructing analytical models of instruction-level execution based on
detailed hardware knowledge and instruction analysis for a single architecture.
Many of these models
predict well for their specific target architecture. For example,
\citet{hong_analytical_2009} present an analytical performance
model for Nvidia GPU architectures that estimates memory-level and thread-level
parallelism. They further extend their model for power prediction \citep{hong_integrated_2010}.
This model achieves a geometric mean error of 13.3\% when
predicting performance of the MERGE \citep{linderman_merge:_2008} benchmarks on
four Tesla generation Nvidia GPUs. It makes extensive use of hardware performance
characteristics, such as timing delays between memory transactions, DRAM access
latency, and instruction execution cycles, and requires an analysis of PTX
assembly instructions. \citet{baghsorkhi_adaptive_2010}
also use deep analytical knowledge of a (single) GPU, and, unlike Hong and Kim,
model branch divergence, bank conflicts, and SIMD pipeline delays.
From the perspective of optimization selection,
\citet{cavazos_automatic_2006} present a probabilistic predictor
of transformation selection using a non-analytical, black-box model based
on an artificial neural network.
\citet{joseph_predictive_2006} use techniques from machine
learning to identify piecewise nonlinearities in cost metrics. Other approaches
emphasize the performance of single subsystems, such as branch
prediction \citep{emer_asim_2002}.

\citet{zhang_quantitative_2011} take a slightly different
approach, using the results of microbenchmarks to derive a throughput model for
instruction pipeline, shared memory, and global memory costs. They focus on
identifying performance bottlenecks and guiding the optimization process
rather than predicting execution time.

Our work differs from previous performance prediction work in five ways:

\begin{itemize}

\item We completely automate the gathering of all performance-relevant kernel
    properties used to model execution time. To our knowledge, this is the first GPU
    performance model to do so.
\item We model execution time without explicit representation of any hardware
    characteristics or behavior.
\item Our model is hardware vendor- and generation-independent, and we
    demonstrate its performance on an AMD GPU and three generations of Nvidia GPUs.
\item Our model is simple and amenable to analysis: through the exposed weights and
  their known meanings, it becomes possible to reason about which parts of the
  kernel execution cost are attributed to which specific operations.
\item Evaluation of the prediction in our model is rapid and simple: Obtaining a cost estimate
    involves only computing a small inner product involving precomputed symbolic expressions
    dependent on problem size parameters.

\end{itemize}

\subsection{Impact on Supercomputing }

Given the shifting of landscape large- and extreme-scale computing in which the
scale of a machine is often determined by power and cooling constraints, it is
inevitable that individual nodes will need to carry a heavier burden than in
prior machine generations. This trend currently shows no signs of
reversing. As a result, more and more complex parallel computing architecture is
found within each individual node. Key to leveraging this within-node parallelism is the
ability to predict its performance on a given computation workload, for needs
such as load balancing, job scheduling, performance optimization, machine design
and qualification, and benchmarking, as detailed in
Section~\ref{sec:applications}. For array-based workloads, as encountered in
much of scientific computing, these needs are met directly by the modeling
machinery supplied by this contribution.

\subsection{Overview of the Contribution}

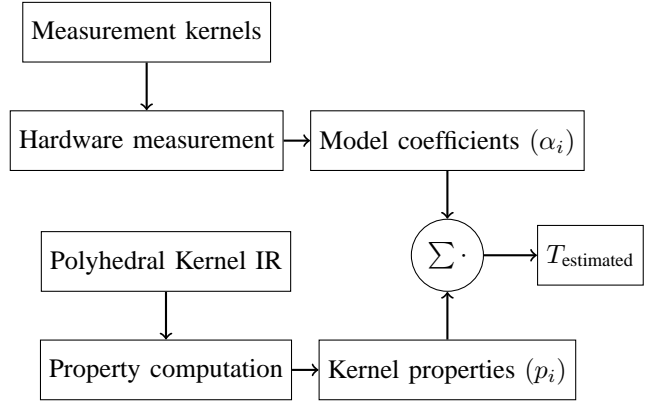
\begin{figure}
  \begin{tikzpicture}
  [
    compbox/.style={draw,minimum height=5ex,text depth=.3ex},
    arr/.style={draw,thick,->}
  ]
    \node [draw,circle] (dot) {$\sum \cdot$};
    \node [compbox,right=2em of dot] (result)
    {$T_{\text{estimated}}$};
    \node [compbox,above=4ex of dot] (coeffs) {Model coefficients
    $(\alpha_i)$};
    \node [compbox,left=1em of coeffs] (measurement) {Hardware measurement};
    \node [compbox,above=4ex of measurement] (measkernel)
    {Measurement kernels};
    \node [compbox,below=4ex of dot] (kprops) {Kernel properties $(p_i)$ };
    \node [compbox,left=1em of kprops] (gather) {Property computation};
    \node [compbox,above=4ex of gather] (poly) {Polyhedral Kernel IR};

    \draw [arr] (measurement) -- (coeffs);
    \draw [arr] (measkernel) -- (measurement);
    \draw [arr] (coeffs) -- (dot);
    \draw [arr] (dot) -- (result);
    \draw [arr] (kprops) -- (dot);
    \draw [arr] (poly) -- (gather);
    \draw [arr] (gather) -- (kprops);
  \end{tikzpicture}
  \caption{Functional overview of the performance model.}
  \label{fig:functional-overview}
\end{figure}

First, in Section~\ref{sec:modeling}, we define a set of kernel properties that
are linearly related to run time. Second, we present a simple procedure for the
extraction of kernel statistics relevant to performance in
Section~\ref{sec:gather}. We further demonstrate how these measures can be used to
obtain the kernel properties that form the basis of our model. Third, in
Section~\ref{sec:weights}, we describe a fitting procedure based on a library of
measurement kernels to determine the parameters of our model for each piece of
hardware on which it is to be used.  Lastly, in Section~\ref{sec:results}, we
evaluate our model's predictive power on a number of GPUs from various hardware
generations and vendors. Figure~\ref{fig:functional-overview} provides a
functional overview of the model.

\section{Modeling Kernel Run Time}
\label{sec:modeling}


We model the execution time of a computational kernel as a linear combination of
individual measures that are chosen a priori to ensure that they contribute to
overall kernel execution time in a linear fashion, i.e.,
\[
T_{\text{wall}}(\mathbf n) \approx \sum_{i=1}^{N_{\text{properties}} }
\alpha_i p_i(\mathbf n),
\]
where $\alpha_i$ is a machine dependent weighting coefficient for the $i$th
contributing cost component, and $p_i(\mathbf n)$ is a kernel-dependent symbolic
expression that, based on size- and loop-bound-related kernel parameters
gathered in the vector $\mathbf n$, accounts for the number of units of the cost
$\alpha_i$ incurred by the given kernel. For kernels with static (i.e., not
data-dependent) control flow, all cost instance expressions $p_i$ are computed
automatically without human intervention. For the kinds of data-dependent
control flow allowed by the transformation system on which we base our modeling
work, a human operator can supply statistics covering typical instances of data
that may be encountered and on which performance is to be modeled. Our model is
fully parametric in the sense that once the symbolic representation of $p_i$ has
been determined from the internal representation of our transformation tool, it
can be cheaply reevaluated for changed values of the parameter vector $\mathbf
n$.

By formulating our model as stated above, we account for costs ultimately
attributable to bandwidth and rate constraints. To retain hardware
independence, we consider modeling the performance effects of
throughput-limiting resource constraints, machine granularities, and latencies
as being out of scope for the current contribution.
Stated another way, we measure the typical rates that the target machine
sustains, and use that as a vehicle for modeling machine performance on other
workloads likely to be bound by the same rate constraint. This specifically
implies that a variety of numbers that are often cited in connection with
performance on GPU hardware are not taken into account in our model.
`Occupancy,' a quantity that describes the fraction of hardware scheduler slots
that the workload may occupy, and thus a measure of the potential for latency
hiding on Nvidia hardware, is one example of a family of effects that our model
does not account for. As a further consequence, our model lacks any
modeling of amortisation or overlapping--any operation is charged at its full,
albeit typical, cost. In part, our overarching goal of achieving hardware
independence necessitates this seemingly draconian restriction in modeling
power. On the other hand, omitting all these effects makes the model an
experiment in simplicity: what quality of results can one obtain in this
setting?

We choose to account for cost components from the following broad categories:
data motion, synchronization, floating point arithmetic, and overhead cost.

\subsection{Cost of Data Motion}
\label{sec:data_motion}

For most types of computational kernels, data motion onto and off the processor
chip is the dominant cost. We account for this in a number of different ways.
At the most basic level, we introduce kernel properties for each access to
global memory, categorized in a number of different ways.

\begin{itemize}
\item The first categorization is performed by the size of the memory access, grouping together
32-bit, 64-bit, and 128-bit accesses.

\item The next categorization occurs by the direction of the memory access,
i.e., load or store.

\item The last categorization occurs by what we call the \emph{amortized stride
fraction}. To determine this fraction, we find the stride, i.e., the address
increment from one abstract SIMD lane (OpenCL work item index) to the next, in
multiples of the size of the overall access. We note that this stride can be
zero, indicating a memory access whose target location does not depend on the
current SIMD lane (`local') index. (In keeping with established terminology, we
term this `uniform access.') The stride forms the denominator of the amortized
stride fraction.

We also find the data utilization ratio on a per-array basis, a quantized
version of which forms the numerator of the amortized stride fraction. This is
accounted for on a per-cell basis, where the size of a `cell' is
given by the data type of the array for which the access is being counted. The
utilization ratio is then determined by counting the total number of cells
being accessed (i.e.,\ the number of accessed cells, as opposed to the total
number of accesses, which can be much larger) and the total number of cells in
the footprint of the access with the `gaps' caused by striding filled in.

For strides 0 and 1, the data utilization ratio is disregarded, and
categorization occurs simply as `stride 0' or `stride 1.' For stride 2, a
utilization ratio of 50\% or less results in a categorization by amortized
stride fraction `1/2,' otherwise we categorize as `2/2.' Counting the `2/2'
access as distinct from the `stride 1' (or `1/1' access)  allows the model to
capture the ability of any caches in the data path to `smooth' out strided
accesses that nonetheless make use of all  accessed data.

We analogously define categories  `1/3,' `2/3,' `3/3,' and `1/4,' `2/4,' `3/4,'
`4/4,' along with `1/$>$4,' `2/$>$4,' `3/$>$4,' `4/$>$4,' accounting for any
accesses with the stride greater than 4.
\end{itemize}

To give the model an opportunity to capture any efficiency gains that are
achievable if both loads and stores are present in the data path, we include a
property for each memory access type containing the minimum between the number
of loads and stores to account for this nonlinearity in the spirit of a
roofline model \citep{williams_roofline_2009}.

Each GPU core (`SM' in Nvidia's terminology, `CU' in AMD's) has its own on-chip local, or `shared,' memory which is
slower than registers, but much faster than global memory. To capture the cost
of moving data from local memory to registers, we define a property for local
loads as well. We do not currently account for stride differences in local
loads.

\subsection{Cost of Floating Point Arithmetic}

While the execution time for many computational kernels is dominated by data
movement, arithmetic operations also contribute to the overall execution
time. The hardware can overlap arithmetic and data movement, and we intend to
account for this in future work, as discussed in Section~\ref{sec:future}.
Currently, our model accounts for the cost of these operations
using several properties related to floating point arithmetic. Since execution
time for arithmetic operations can be affected by both the operation kind and
the data type of the operands, we separate kernel properties relating to
operations by both kind and operand data type. Our operation kind categories
include:

\begin{itemize}
    \item Addition and subtraction
    \item Multiplication
    \item Division
    \item Exponentiation
    \item Other special functions
\end{itemize}

For each of these operation kinds, our model includes one property of total
operation counts for 32-bit and 64-bit floating point operand data
types. Integer arithmetic is not accounted for in this version of the model
because it is not typically a dominant contributor to computational cost for the
kernels targeted by our model, and code involving integer arithmetic is often heavily optimized by
modern compilers.

\subsection{Cost of Synchronization}

Barriers in GPU kernels stop execution of every thread within a work group
until all threads have reached the barrier. Thus, thread synchronization can be
a significant contributor to execution time. To account for this within-group
thread synchronization costs we include a property containing the total number
of barriers encountered by all threads.

\subsection{Overhead }

Launching any kernel, regardless of complexity, incurs a constant overhead cost.
Additionally, our experiments running empty kernels revealed that launch
overhead increases with the number of work groups launched. We account for
these costs in our model with two properties. The first is a constant property
(i.e., 1), which accounts for the portion of the launch overhead that remains
constant. The second is the total work group count, which accounts for the
overhead that increases with the number of work groups.


\section{Gathering Kernel Statistics}
\label{sec:gather}


In this section, we describe a methodology to automatically gather data to help
determine the kernel properties used by our modeling process. To do so, we
leverage the \loopy~\citep{kloeckner_loopy_2014,kloeckner_loopy_2015}
programming system in a number of ways:
\begin{itemize}
\item we express our kernels in its intermediate representation,
\item we use its transformation vocabulary to obtain numerous
  computationally different but mathematically equivalent variants for our
  measurements,
\item its code generation capability supplies an executable (OpenCL) version of
  the code which we use to carry out our measurements,
\item and finally, we make use of \loopy's polyhedrally-based internal
  representation to support the automatic extraction of kernel properties.
\end{itemize}
We note that this piece of our work is notionally independent of our model in
the sense that, while it is convenient to have the ability to automatically
extract the properties being used as part of the model, it is not technically
necessary and could be achieved either by hand or in a technologically
different manner. It is relevant to the present discussion insofar as
it confirms that the properties \emph{can} be determined in an automated
fashion.

\subsection{\loopy}

\loopy\ is a programming system
for array computations that targets CPUs, GPUs, and other, potentially
heterogeneous, compute architectures. It is based on the idea that the
mathematical intent and the computational minutiae of a computation should be
strictly separated. To attain that goal, \loopy\ realizes programs as objects in
a host programming language (Python in this concrete case) that can be
manipulated from their initial, ``clean,'' mathematical statement into highly
device-specific, optimized versions via a broad array of transformations.

We briefly examine \loopy's model of a program (or `kernel').  A
very simple
example shall serve as an introduction.  This kernel
reads in one vector, doubles it, and writes the result to another:
\medskip

\begin{lstlisting}[style=custompython,gobble=2]
  knl = loopy.make_kernel(
      "{[i]: 0<=i<n}",  # loop domain
      "out[i] = 2*a[i]")  # instructions
\end{lstlisting}

\medskip
\noindent
The above snippet of code illustrates the main components of the internal representation:
\begin{itemize}
  \item The \emph{loop domain}: \verb|{ [i]: 0<=i<n }|. This defines
    the integer values of the loop variables for which instructions
    (see below) will be executed.
    It is written in the syntax of the \texttt{isl} library
    \citep{verdoolaege_isl_2010}.

    To accommodate data-dependent control flow, a \emph{tree of loop domains}
    is permitted, allowing more deeply nested domains to depend on data
    fetched by instructions executed through loop domains closer to the root.

  \item The \emph{instructions} to be executed: \verb|out[i] = 2*a[i]|. These are scalar
  assignments between array elements, consisting of a left-hand
  side assignee and a right-hand side expression.
  Right-hand side expressions are allowed to contain the usual mathematical
  operators, and function calls.

  Each instruction is executed once for each integer point in the projection of
  the loop domain onto its relevant set of loop variables.

  To facilitate ordering, \loopy\ allows the specification of a directed
  acyclic graph of dependencies in which the instructions form the nodes, and the
  dependency annotations from the edges.
\end{itemize}

\subsection{Extracting Kernel Statistics}
\label{section:stats}

The basic mathematical primitive underpinning our data gathering strategy is
the ability to count the number of integer points in a subset of the
$d$-dimensional integer tuples $\mathbb Z^d$ specified by affine inequalities
connected in disjunctive normal form (i.e., a disjunction of conjunctions of
affine inequalities). The output of this operation is a piecewise
quasi-polynomial in terms of size parameters that may occur as part of the
specification of the set of integers. We make use of \texttt{barvinok} library
in conjunction with the \texttt{isl} library
\citep{verdoolaege_counting_2007,verdoolaege_isl_2010} to perform this
operation, with a fallback to a less accurate, simpler counting technique
that is used should \texttt{barvinok} not be available. \texttt{barvinok}
in turn is based on Barvinok's algorithm \citep{barvinok1994polynomial}.

To obtain a count of, say, the number of memory references of a certain kind,
we proceed as follows:

\begin{algorithm}
  \caption{Determine per-kernel count of per-instruction property}
  \begin{algorithmic}
    \FOR{each instruction $i$ in the kernel}
      \STATE Compute the projection $\pi_i(D_i)$ of the loop domain $D_i$
        onto the relevant set of loop indices.
      \STATE Obtain a symbolic count $|\pi_i(D_i)|$ of the number of integer
      points in the projection (representing the number of times that instruction
      will be executed).
      \STATE Next, process the instruction to find the number of desired
      operations $n_{\text{ops},i}$ (e.g.\ by traversing the left- and
      right-hand-side expressions).
    \ENDFOR
    \STATE
    Find the overall count of the desired operations as
    \begin{equation}
      n_{\text{ops}} = \sum_{\text{Instruction $i$}} |\pi_i(D_i)| \cdot n_{\text{ops}, i}.
      \label{eq:opcount}
    \end{equation}
  \end{algorithmic}
\end{algorithm}
Some counting operations require ancillary processing. For instance, determining
the number of floating point operations of a certain type requires knowing the
result type, which is provided by a type inference pass.
Practically speaking, many types of counts are extracted at once and maintained
in a mapping, with sufficiently specific keys to supply detailed data for
the computation of kernel properties, and with values of piecewise quasi-polynomials.
All arithmetic in \eqref{eq:opcount} is then carried through to the values
of the mapping and performed on the piecewise quasi-polynomials
therein.

To determine the amortized stride fraction, the overall size of the accessed footprint
is needed. This is found as follows:

\begin{algorithm}
  \caption{Determine accessed index footprint $F_v\subset \mathbb
  Z^{d'}$ for variable $v$}
  \begin{algorithmic}
    \STATE Let $v$ be a $d'$-dimensional array
    \FOR{each instruction $i$ in the kernel}
      \STATE Compute the projection $\pi_i(D_i)$ of the loop domain $D_i$
        onto the relevant set of loop indices.
      \FOR{each access $j$ to $v$ in instruction $i$}
        \STATE Determine the
          multi-dimensional index mapping $I_j:\mathbb Z^{d}\to \mathbb N_0^{d'}$ that takes a tuple
          of loop variables to the accessed indices. For example, the access \verb|a[2*i+1, j+1]|
          would have an index mapping of $I_j(i,j)=(2i-1, j+1)$.
      \ENDFOR
    \ENDFOR
    \STATE
    Find the overall accessed footprint as
    \begin{equation*}
      F_v= \bigcup_{\text{Instruction $i$, access $j$}} I_j(\pi_i(D_i)).
    \end{equation*}
  \end{algorithmic}
\end{algorithm}

To determine the accessed stride fraction, we obtain a count of the number of
integer points of $F_v$ as well as those of its filled-in counterpart with any
axis-0 striding removed. By taking a ratio of the two, we find the
stride fraction.

We count loads from local memory just as we do global memory transactions,
although strides and array names are not tracked as part of the summation mapping.

Counting barrier synchronizations requires yet another approach, as these are
not apparent in \loopy\ code without a \emph{schedule}. The schedule is found
automatically by a search procedure and determines the ordering of instructions
and the nesting of loops as well as the location and nesting of required
barrier synchronizations. Once a schedule is obtained, the counting process
proceeds much as above, using the schedule information to obtain the relevant
set of loop indices on which to project.


\section{Fitting Model Weights}
\label{sec:weights}


\subsection{Constructing Kernel Measurement Set}

We fit the model weights to a particular GPU according to data gathered from
example kernel execution. For this purpose, we have built a set of measurement
kernels to provide the model with `informative' examples. Some of the
measurement kernels represent common basic computations, including matrix
multiplication, matrix transposition, and vector operations. Others are designed
to directly exercise one or more operations that are being captured by our
property extraction. To produce the results presented in this paper, we used 9
classes of measurement kernels. Six thread group size sets are referenced in the
measurement kernel list below and again later in the test kernel list:
\begin{itemize}
    \item 1-D Small: $\{(192\times 1), (224\times 1), (256\times 1)\}$
    \item 1-D Med: $\{(128\times 1), (256\times 1), (384\times 1)\}$
    \item 1-D Large: $\{(256\times 1), (384\times 1), (512\times 1)\}$
    \item 2-D Small: $\{(16\times 12), (16\times 14), (16\times 16)\}$
    \item 2-D Med: $\{(16\times 12), (16\times 16), (32\times 16)\}$
    \item 2-D Large: $\{(16\times 16), (24\times 16), (32\times 16)\}$
\end{itemize}
We use the following measurement kernel classes:
{%
\renewcommand\labelitemii{\tiny$\bullet$}
\def\measureknlitem#1{\item \emph{#1}. }
\begin{itemize}
    \measureknlitem{Matrix Multiplication}
    Performs a tiled multiplication of two matrices of size $n\times m$ and
    $m\times l$ (row-major data layout). Prefetches $\gsize\times \gsize$ tiles
    into local memory. Shape cases:
    \begin{itemize}
        \item $n=m=l$
        \item $n=m$ and $l=n/2$
        \item $n=l$ and $m=n/2$
        \item $m=l$ and $n=m/2$
    \end{itemize}
    Four size cases: $n=2^{p+t}$ where $t=0,1,2,3$. For each GPU we choose $p\in[7,8,9]$
    depending on launch overhead and memory limitations.
    Group sizes:
    \begin{itemize}
        \item R9 Fury: 2-D Small
        \item Tesla C2070, K40c: 2-D Med
        \item Titan X: 2-D Large
    \end{itemize}

    \measureknlitem{Naive Matrix Multiplication}
    Performs a non-tiled multiplication of two square $n\times n$ matrices
    (row-major data layout) with each thread computing one element of the result
    as the inner product of the corresponding row and column. Four size cases:
    $n=2^{p+t}$ where $t=0,1,2,3$. For each GPU we choose $p\in[6,8,9]$ depending on launch
    overhead and memory limitations.
    Group sizes:
    \begin{itemize}
        \item R9 Fury: 2-D Small
        \item Tesla C2070, K40c: 2-D Med
        \item Titan X: 2-D Large
    \end{itemize}

    \measureknlitem{Vector Scale and Add}
    Multiplies two $n\times 1$ vectors each by a scalar and adds the result.
    Each thread computes one value in the result. Three stride configurations:
    \begin{enumerate}
        \item Operates on every element, resulting in stride-1 access pattern.
        \item Operates on every other element, resulting in stride-2 access
            pattern.
        \item Operates on every third element, resulting in stride-3 access
            pattern.
    \end{enumerate}
    Four size cases: $n=2^{p+2t}$ where $t=0,1,2,3$. For each GPU we choose $p\in[18,20,21]$
    depending on launch overhead and memory limitations.
    Group sizes:
    \begin{itemize}
        \item R9 Fury: 1-D Small
        \item Titan X, Tesla C2070, K40c: 1-D Large
    \end{itemize}

    \measureknlitem{Transpose}
    Performs a transpose operation on a square $n\times n$ matrix (row-major
    data layout), storing the result in a second matrix. Each thread moves one
    matrix element.
    Three prefetch/stride configurations:
    \begin{enumerate}
        \item Prefetches $(\gsize\times \gsize)$ tiles into local memory to allow
            stride-1 access pattern for reads and writes.
        \item Does not prefetch; stride-1 access pattern for writes but not
            reads.
        \item Does not prefetch; stride-1 access pattern for reads but not
            writes.
    \end{enumerate}
    Four size cases: $n=2^{p+t}$ where $t=0,1,2,3$. For each GPU we choose $p\in[10,11]$
    depending on launch overhead and memory limitations.
    Group sizes:
    \begin{itemize}
        \item R9 Fury: 2-D Small
        \item Titan X, Tesla C2070, K40c: 2-D Med
    \end{itemize}

    \measureknlitem{Stride-1 Global Access}
    Copies $n\times 1$ arrays or sums of arrays. Three configurations:
    \begin{enumerate}
        \item Copy one array (1 load, 1 store)
        \item Add 4 arrays, store the result in a 5th array (4 loads, 1
            store)
        \item Store the index of each element into one array (0 loads, 1
            store)
    \end{enumerate}
    Nine size cases: $n=2^{p+t}$ where $t=0,1,\dotsc,8$. For each GPU we choose $p\in[17,18,19,20]$
    depending on launch overhead and memory limitations.
    Group sizes:
    \begin{itemize}
        \item R9 Fury: 1-D Small
        \item Tesla C2070, K40c: 1-D Med
        \item Titan X: 1-D Large
    \end{itemize}

    \measureknlitem{Stride-2 Filled Global Access}
    Produces summations of pairwise sums of consecutive elements in a $2\times
    n$ array (column-major data layout), prefetching elements in a stride-2
    access pattern (i.e., first fetches elements $i, i+2, i+4, \dotsc,
    i+2(\gsize-1)$, then fetches elements $i+1, i+3, i+5, \dotsc, i+1+2(\gsize-1)$
    then adds pairs). Each of $n$ threads performs a summation over 256 of these
    pairwise sums and stores the result in one index of a $1\times n$ output
    array. Four size cases: $n=2^{p+3t}$ where $t=0,1,2,3$. For each GPU we
    choose $p\in[15,16,17]$ depending on launch overhead and memory limitations.
    Group sizes:
    \begin{itemize}
        \item R9 Fury: 1-D Small
        \item Tesla C2070, K40c: 1-D Med
        \item Titan X: 1-D Large
    \end{itemize}

    \measureknlitem{Stride-3 Filled Global Access}
    Same as stride-2 filled access measurement kernels above, but produces
    triowise sums on a $3\times n$ array, producing a stride-3 access pattern.

    \measureknlitem{Arithmetic Operations}
    Performs arithmetic operations without reading global memory data, storing
    results in an $n\times n$ output matrix (row-major data layout). Each thread
    produces one element in the output matrix as a summation over $k$ indices of
    an expression involving 6-10 arithmetic operations of a specific type using the
    index.
    Arithmetic types (separate kernel for each):
    \begin{itemize}
        \item{Addition and subtraction}
        \item{Multiplication}
        \item{Division}
        \item{Exponentiation}
        \item{Rsqrt function (since this appears in our test kernels)}
    \end{itemize}
    Nine size cases: $k$ takes on values $256,512,728$, and for each value of $k$, $n$
    takes on three values of the form $2^{p+t}$ where $t=0,1,2$. For each GPU we
    choose $p\in[7,8]$ depending on launch overhead and memory limitations.
    Group sizes:
    \begin{itemize}
        \item R9 Fury: 2-D Small
        \item Tesla C2070, K40c: 2-D Med
        \item Titan X: 2-D Large
    \end{itemize}

    \measureknlitem{Empty kernel}
    Performs no operations or memory accesses. Launches thread groups as if each
    thread operates on one element in a non-existent $n\times n$ matrix. Six size
    cases: $n=2^{p+t}$ where $t=\{0,1,2,3,4,5\}$. For each GPU we choose $p\in[8,9,10]$ depending on
    launch overhead and memory limitations.
    Group sizes:
    \begin{itemize}
        \item R9 Fury: 2-D Small
        \item Tesla C2070, K40c: 2-D Med
        \item Titan X: 2-D Large
    \end{itemize}

\end{itemize}
}

\subsection{Measurement Kernel Execution}

To facilitate organization and execution of our measurement kernels, we have
constructed infrastructure to house collections of kernels and associated
optimization configurations. This mechanism transforms kernels according to
their configuration lists, launches each configuration, and saves the data
gathered for future use. Kernels are compiled to and executed using OpenCL.

We intend to make our set of benchmark kernels (both those used for measurement
as well as gathering of results) available for other researchers to use, and
also to facilitate easy reproduction of our results. For the moment however,
this remains future work.

Consistent timing of measurement kernel execution is crucial to the accuracy of
our model, and we found that consistency decreased significantly as small
execution times approached the kernel launch overhead. This overhead varied
between GPUs, with the AMD GPU having the highest launch overhead. For this
reason, on each GPU we first run the empty kernel to determine launch overhead,
and then set the minimum size configuration for each measurement and testing
kernel to meet or exceed this run time.

We then time 30 runs of each kernel. The arrays are allocated on first-touch,
which produces a greater execution time for the first run. Additionally, our
experiments revealed that the second of the 30 runs varies more than the rest.
Thus, we disregard the first 4 runs and take the minimum of the remaining
execution times. Taking the average also produced consistent results, and we
found that the minimum differed from the average by less than 5\% when execution
times significantly exceeded the launch overhead.

After timing each kernel, we gather the kernel statistics as described in
Section~\ref{section:stats} and form the properties described in
Section~\ref{sec:modeling}.

\subsection{Calculating the Weights}

After running the measurement kernel set and forming the properties for each
kernel, we produce a property matrix with one row per kernel and one column per
property. Since we would like to minimize relative error instead of absolute
error, we next divide each property by the observed run time for the
corresponding measurement kernel. We find the weights $(\alpha_i)$ as the ones
that minimize
\[
\sum_{j=1}^{N_\text{cases}}
\bigg(1
  - \frac{
      \sum_{i=1}^{N_{\text{properties}} } \alpha_i p_{i,\text{test $j$}}(\mathbf n_j)\
     }{
      T_{\text{wall, test $j$, measured}}
     }
\bigg)^2.
\]


\section{Results}
\label{sec:results}

We demonstrate the predictive accuracy of our model on four GPUs:
\begin{itemize}
    \item Nvidia GTX Titan X (Maxwell generation)
    \item Nvidia Tesla K40 (Kepler generation)
    \item Nvidia Tesla C2070 (Fermi generation)
    \item AMD Radeon R9 Fury
\end{itemize}
For each GPU, we run four test kernels:
{%
\def\testknlitem#1{\item \emph{#1}. }
\begin{itemize}
    \testknlitem{Finite Differences}
    Applies a 5-point stencil with a quadratic source term on a square $n\times
    n$ matrix (row-major data layout), prefetching $\gsize\times \gsize$ tiles
    into local memory, plus halo elements. Four size cases: $n=2^{p+t}$ where
    $t=0,1,2,3$. For each GPU we choose $p\in[10,11]$ depending on launch overhead and memory
    limitations.
    Group and problem sizes:
    \begin{itemize}
        \item R9 Fury: 2-D Small, $p=10$
        \item Tesla C2070: 2-D Med, $p=10$
        \item Tesla K40c: 2-D Med, $p=11$
        \item Titan X: 2-D Large, $p=11$
    \end{itemize}

    \testknlitem{`Skinny' Matrix Multiplication}
    Performs a tiled multiplication of two matrices of size $n\times m$ and
    $m\times l$, with $n=l=m/8$ (row-major data layout). Prefetches $\gsize\times
    \gsize$ tiles into local memory.
    Four size cases: $n=2^{p+t}$ where $t=0,1,2,3$. For each GPU we choose $p\in[9,10]$ depending on
    launch overhead and memory limitations.
    Group and problem sizes:
    \begin{itemize}
        \item R9 Fury: 2-D Small, $p=9$
        \item Tesla C2070, K40c: 2-D Med, $p=9$
        \item Titan X: 2-D Large, $p=10$
    \end{itemize}

    \testknlitem{Convolution}
    Applies three $7\times 7$ image filters to three $n\times n$ RGB images,
    i.e.,
    it computes
    \begin{multline*}
            r_{i,j,x,y} = \\ \sum_{-w\le\xi, \eta\le w,c=0,1,2}
            m_{i,w+x-\xi, w+y-\eta, c}\\
            \cdot f_{j, w+\xi, w+\eta, c},
    \end{multline*}
    where $f_j$ indicates the $j$th filter image ($j=0,1,2$), and $w=3$ denotes the
    positive/negative index range (along both axes) of all filters.
    Four size cases: $n=2^{p+t}$ where $t=0,1,2,3$. For each GPU we choose $p\in[6,7,8]$ depending on
    launch overhead and memory limitations.
    Group and problem sizes:
    \begin{itemize}
        \item R9 Fury: 2-D Small, $p=7$
        \item Tesla C2070: 2-D Med, $p=6$
        \item Tesla K40c: 2-D Med, $p=7$
        \item Titan X: 2-D Large, $p=8$
    \end{itemize}

    \testknlitem{N-Body}
    Given a $3\times n$ array of $n$ positions (column-major data layout),
    computes the sum of the inverses of the distances between each position and
    every other position, prefetching position data in $3\times \gsize$ blocks.
    Each thread computes said sum for one position.
    Four size cases: $n=2^{p+t}$ where $t=0,1,2,3$. For each GPU we choose $p\in[10,11]$ depending on
    launch overhead and memory limitations.
    Group and problem sizes:
    \begin{itemize}
        \item R9 Fury: 1-D Small, $p=10$
        \item Tesla C2070, K40c: 1-D Med, $p=11$
        \item Titan X: 1-D Large, $p=11$
    \end{itemize}

\end{itemize}

As discussed in Section~\ref{sec:modeling}, we do not account for occupancy in
our model. Thus, aside from the overhead in launching extra thread work groups,
which we account for and can be significant, our model produces the same
prediction for kernel configurations that differ only in work group size unless
the work group size affects the kernel properties in some way (e.g., the number
of memory transactions may be affected by work group size if work groups require
data from halo elements). To obtain a representative sample of performance
across thread group sizes, we run each of our measurement kernels with three
different thread group sizes ranging from 128 to 512 threads, as described
above. These configurations vary depending on the kernel and hardware
limitations; the Radeon R9 Fury limits group sizes to 256. Most of the
measurement kernels yield full occupancy on all 3 Nvidia GPUs, and we report
results for test kernels with 256-thread groups that yield full occupancy.
Run-time generally varied by less than 30\% due to thread group size changes.
Among thread work group sizes, model predictions reported for the 256-thread
work groups were neither the most accurate nor the least accurate.

\begin{table*}
    \begin{center}
        \begin{tabular}{lrrcrrcrrcrrr}
  \toprule
 & \multicolumn{2}{c}{Nvidia} && \multicolumn{2}{c}{Nvidia} &&
  \multicolumn{2}{c}{Nvidia} && \multicolumn{2}{c}{AMD} & Cross-GPU \\
  Kernel & \multicolumn{2}{c}{GTX Titan X} && \multicolumn{2}{c}{Tesla C2070} &&
  \multicolumn{2}{c}{Tesla K40} && \multicolumn{2}{c}{Radeon R9 Fury} &
  Geometric Mean \\
  \midrule
  Finite Difference & \multicolumn{2}{c}{\textbf{0.30}} && \multicolumn{2}{c}{\textbf{0.10}} && \multicolumn{2}{c}{\textbf{0.01}} && \multicolumn{2}{c}{\textbf{0.63}} & \textbf{0.11} \\
  \cmidrule{2-3} \cmidrule{5-6} \cmidrule{8-9} \cmidrule{11-12}
  a. & 0.32 & 0.41 && 0.44 & 0.40 && 0.70 & 0.70 && 0.16 & 0.22 & \\
  b. & 1.03 & 1.39 && 1.35 & 1.21 && 2.37 & 2.42 && 0.27 & 0.48 & \\
  c. & 4.27 & 5.32 && 4.98 & 4.46 && 9.17 & 9.31 && 0.89 & 1.55 & \\
  d. & 15.33 & 21.05 && 19.55 & 17.43 && 37.34 & 36.87 && 3.23 & 5.81 & \\
  \addlinespace[2pt]
  \midrule
  Skinny MM & \multicolumn{2}{c}{\textbf{0.08}} && \multicolumn{2}{c}{\textbf{0.10}} && \multicolumn{2}{c}{\textbf{0.13}} && \multicolumn{2}{c}{\textbf{0.28}} & \textbf{0.13} \\
  \cmidrule{2-3} \cmidrule{5-6} \cmidrule{8-9} \cmidrule{11-12}
  a. & 0.18 & 0.14 && 0.28 & 0.18 && 0.27 & 0.14 && 0.25 & 0.14 & \\
  b. & 0.56 & 0.55 && 0.58 & 0.51 && 0.41 & 0.28 && 0.41 & 0.18 & \\
  c. & 3.52 & 3.81 && 3.35 & 3.16 && 1.65 & 1.33 && 0.87 & 0.55 & \\
  d. & 27.23 & 29.73 && 23.26 & 24.23 && 9.62 & 9.71 && 3.23 & 3.44 & \\
  \addlinespace[2pt]
  \midrule
  N-Body & \multicolumn{2}{c}{\textbf{0.32}} && \multicolumn{2}{c}{\textbf{0.27}} && \multicolumn{2}{c}{\textbf{0.54}} && \multicolumn{2}{c}{\textbf{0.76}} & \textbf{0.43} \\
  \cmidrule{2-3} \cmidrule{5-6} \cmidrule{8-9} \cmidrule{11-12}
  a. & 0.48 & 0.16 && 1.06 & 0.48 && 0.99 & 0.24 && 0.39 & 0.14 & \\
  b. & 0.90 & 0.38 && 2.67 & 1.51 && 1.99 & 0.59 && 0.64 & 0.15 & \\
  c. & 1.83 & 1.29 && 7.41 & 5.66 && 4.26 & 2.01 && 1.31 & 0.22 & \\
  d. & 4.49 & 4.90 && 24.58 & 22.26 && 10.90 & 7.66 && 2.32 & 0.49 & \\
  \addlinespace[2pt]
  \midrule
  Convolution & \multicolumn{2}{c}{\textbf{0.10}} && \multicolumn{2}{c}{\textbf{0.13}} && \multicolumn{2}{c}{\textbf{0.03}} && \multicolumn{2}{c}{\textbf{0.23}} & \textbf{0.10} \\
  \cmidrule{2-3} \cmidrule{5-6} \cmidrule{8-9} \cmidrule{11-12}
  a. & 0.49 & 0.47 && 0.34 & 0.25 && 0.43 & 0.33 && 0.28 & 0.19 & \\
  b. & 1.54 & 1.64 && 0.62 & 0.60 && 1.08 & 0.96 && 0.43 & 0.35 & \\
  c. & 5.73 & 6.32 && 1.73 & 2.01 && 3.49 & 3.48 && 1.13 & 1.02 & \\
  d. & 19.32 & 25.04 && 6.19 & 7.65 && 13.30 & 13.56 && 6.75 & 3.69 & \\
  \addlinespace[2pt]
  \midrule
  Cross-Kernel &&&&&&&&&&&& \\
  Geometric Mean & \multicolumn{2}{c}{\textbf{0.16}} && \multicolumn{2}{c}{\textbf{0.14}} && \multicolumn{2}{c}{\textbf{0.06}} && \multicolumn{2}{c}{\textbf{0.42}} & \\
  \bottomrule
\end{tabular}

        \medskip
        \caption{Predicted vs.\ actual execution times (ms) for test kernels, and
        geometric mean of relative error.}
        \label{table:results}
    \end{center}
\end{table*}

Table~\ref{table:results} displays predicted and actual execution
times in milliseconds for each test kernel on each GPU\@. We measure
model error as the ratio of absolute value of the difference between predicted and
actual execution times and the actual execution time. Since these are
normalized values, we summarize them using the geometric mean for
reasons laid out by \citet{fleming_how_1986}.

The geometric means of relative absolute error across all test kernels on the
Nvidia GTX Titan X, Tesla C2070, and Nvidia Tesla K40c were 16\%, 14\%, and 6\%,
respectively. Performance on the Radeon was found to be irregular and, as such,
less amenable to being captured by our model. Even so, it predicts two of the
kernels reasonably well; the geometric mean errors for the `skinny' matrix
multiplication and convolution kernels on the were 28\% and 23\%, respectively.

Across GPUs, our model predicts the finite difference, skinny matrix
multiplication, and convolution kernels with mean errors of less than 13\%. It
had more difficulty predicting the N-Body kernel on all GPUs, yielding a mean
error of 43\%.

The measurement kernels described in Section~\ref{sec:weights} do not contain
instances of every model property subclass described in
Section~\ref{sec:modeling}; they contain instances of every property relevant to
the test kernels. Example weights for these properties produced for the Radeon
R9 Fury are displayed in Table~\ref{table:weights}.
It is worth noting that the weights determined by our fitting
procedure carry units of seconds per operation and are amenable to
direct interpretation. Beyond that, they allow direct conclusions
about sustained typical rates for different types of hardware and
are directly comparable across devices.

\begin{table}
    \begin{center}
        \begin{tabular}{lr}
  \toprule
  Property & Weight \\
  \midrule
  Addition/Subtraction & \ 6.81e-13 \\
  Multiplication & \ 5.68e-13 \\
  Exponentiation & \ 3.91e-13 \\
  Other\ Ops (rsqrt) & \ 1.61e-12 \\
  Local\ F32\ Loads & -1.76e-12 \\
  F32\ Stride-1\ Loads & \ 8.27e-12 \\
  F32\ Stride-2\ (100\%)\ Loads & \ 9.82e-13 \\
  F32\ Stride-3\ ( 33\%)\ Loads & \ 2.89e-11 \\
  F32\ Stride-3\ (100\%)\ Loads & \ 9.30e-13 \\
  F32\ Uncoalesced\ (100\%)\ Loads & \ 2.67e-12 \\
  F32\ Stride-1\ Stores & \ 6.52e-12 \\
  F32\ Uncoalesced\ (100\%)\ Stores & \ 3.55e-10 \\
  Min(Stride-1\ Loads, Stride-1\ Stores) & -6.63e-12 \\
  Barriers & \ 4.26e-11 \\
  Thread\ Groups & \ 3.75e-09 \\
  Const(1) & \ 1.29e-04 \\
  \bottomrule
\end{tabular}

        \medskip
        \caption{Example set of property weights (in units of seconds per
        operation) produced for the AMD Radeon R9 Fury. Percentages represent
    the utilization ratio discussed in Section~\ref{sec:data_motion}}
        \label{table:weights}
    \end{center}
\end{table}


\section{Conclusions}
\subsection{Potential Applications}
\label{sec:applications}

Being able to reason and make predictions about the wall time cost of a given
computation is a foundational capability for a large number of activities in
high-performance computing:

\begin{itemize}

\item In \emph{performance optimization}, it can aid an optimization tool in exploring
a search space of program transformations.

\item In \emph{algorithm design}, it can provide guidance on which aspects of the
workload under consideration are the biggest contributors to computational cost.

\item In \emph{load balancing}, accurate predictions of workload run times
enable better scheduling decisions, thereby facilitating the reduction of idle
time and making better use of available computational resources. This need is
particularly salient when a workload is to be moved across heterogeneous compute
resources.

\item In \emph{machine bringup and qualification}, our measurement procedure can
expose bottlenecks as well as unexpected interactions and help enable
comparisons between different processor architectures.

\end{itemize}

Independence of particular architecture as well as rapid model evaluation make
our methodology particularly suitable for these application scenarios.

\subsection{Future Work}
\label{sec:future}

Our work permits a number of immediate extensions. Perhaps the most obvious one
of these would be to investigate how much information on potentially overlapped
operations can be obtained through a fitted model. A prominent example of this
would be overlapping arithmetic with data motion. Another possible extension
would handle resource limitations on, say, the number of registers, the
amount of local memory, and their respective effects on performance.

Other aspects of GPU execution cost may be simpler to account for. For example,
bank conflicts in local memory may be handled by binning the stride of local
memory access.

Another interesting extension would be to study our model's ability to select
the optimal set of kernel configurations (i.e., the set that produces the
fastest kernel) from a collection of potential optimizations. This ability,
combined with the rapid evaluation speed of our model, would enable runtime
performance tuning of GPU kernels.

Another immediate extension of this work would be to examine its applicability
on CPU-type architectures. For these types of machines, data motion cost would
necessarily need to include a model of cache and data reuse. It would also be
interesting to investigate to what extent a version of this model can apply to
current and future  wide-vector manycore accelerators of the Xeon Phi and
related families.

\subsection{Summary}

This paper makes the following contributions:
\begin{itemize}
\item
  It identifies a set of hardware-independent kernel properties that suffice to
  account for kernel run times with considerable accuracy.
\item
  It describes a procedure for the automatic extraction of symbolic counts
  from the internal representation of our transformation engine, based
  on the polyhedral model. The representation of these counts as piecewise
  quasi-polynomials offers both efficient evaluation and considerable
  generality.
\item
  It describes a set of measurements as well as a fitting procedure to, once
  again in a black box and unassisted fashion, determine hardware- specific
  weights for each of the properties determined above.
\end{itemize}

We have demonstrated an alternative to previous GPU performance models that can
be easily fitted to new hardware and allows rapid, runtime performance
prediction. This speed and versatility turns out to require minor if any sacrifices in
prediction accuracy compared to models in the literature. To our knowledge, this is the first GPU performance model that
collects all performance-relevant information automatically and utilizes no
explicit knowledge of hardware characteristics.


\section*{Acknowledgments}
\ifthenelse{\boolean{doubleblind}}{
  (omitted in accordance with double-blind review guidelines)
}{
  The authors' work was supported in part by US Navy ONR grant numbers
  N00014-14-1-0117, and by the National Science Foundation under grant
  numbers DMS-1418961 and CCF-1524433. AK also gratefully acknowledges a hardware
  gift from Nvidia Corporation.

  Opinions expressed herein are those of the authors and in no way reflect the
  official position of any of the funding agencies.
}

\clearpage
\bibliography{refs}{}

\begin{thebibliography}{18}
\providecommand{\natexlab}[1]{#1}
\providecommand{\url}[1]{\texttt{#1}}
\expandafter\ifx\csname urlstyle\endcsname\relax
  \providecommand{\doi}[1]{doi: #1}\else
  \providecommand{\doi}{doi: \begingroup \urlstyle{rm}\Url}\fi

\bibitem[Baghsorkhi et~al.(2010)Baghsorkhi, Delahaye, Patel, Gropp, and
  Hwu]{baghsorkhi_adaptive_2010}
S.~S. Baghsorkhi, M.~Delahaye, S.~J. Patel, W.~D. Gropp, and W.-m.~W. Hwu.
\newblock An {Adaptive} {Performance} {Modeling} {Tool} for {GPU}
  {Architectures}.
\newblock In \emph{Proceedings of the 15th {ACM} {SIGPLAN} {Symposium} on
  {Principles} and {Practice} of {Parallel} {Programming}}, {PPoPP} '10, pages
  105--114, New York, NY, USA, 2010. ACM.
\newblock ISBN 978-1-60558-877-3.
\newblock \doi{10.1145/1693453.1693470}.

\bibitem[Barvinok(1994)]{barvinok1994polynomial}
A.~I. Barvinok.
\newblock A polynomial time algorithm for counting integral points in polyhedra
  when the dimension is fixed.
\newblock \emph{Mathematics of Operations Research}, 19\penalty0 (4):\penalty0
  769--779, 1994.

\bibitem[Cavazos et~al.(2006)Cavazos, Dubach, Agakov, Bonilla, O'Boyle, Fursin,
  and Temam]{cavazos_automatic_2006}
J.~Cavazos, C.~Dubach, F.~Agakov, E.~Bonilla, M.~F. O'Boyle, G.~Fursin, and
  O.~Temam.
\newblock Automatic performance model construction for the fast software
  exploration of new hardware designs.
\newblock In \emph{Proceedings of the 2006 international conference on
  {Compilers}, architecture and synthesis for embedded systems}, pages 24--34.
  ACM, 2006.

\bibitem[Emer et~al.(2002)Emer, Ahuja, Borch, Klauser, Luk, Manne, Mukherjee,
  Patil, Wallace, Binkert, and {others}]{emer_asim_2002}
J.~Emer, P.~Ahuja, E.~Borch, A.~Klauser, C.-K. Luk, S.~Manne, S.~S. Mukherjee,
  H.~Patil, S.~Wallace, N.~Binkert, and {others}.
\newblock Asim: {A} performance model framework.
\newblock \emph{Computer}, 35\penalty0 (2):\penalty0 68--76, 2002.

\bibitem[Fleming and Wallace(1986)]{fleming_how_1986}
P.~J. Fleming and J.~J. Wallace.
\newblock How {Not} to {Lie} with {Statistics}: {The} {Correct} {Way} to
  {Summarize} {Benchmark} {Results}.
\newblock \emph{Commun. ACM}, 29\penalty0 (3):\penalty0 218--221, Mar. 1986.
\newblock ISSN 0001-0782.
\newblock \doi{10.1145/5666.5673}.

\bibitem[Hong and Kim(2009)]{hong_analytical_2009}
S.~Hong and H.~Kim.
\newblock An {Analytical} {Model} for a {GPU} {Architecture} with
  {Memory}-level and {Thread}-level {Parallelism} {Awareness}.
\newblock In \emph{Proceedings of the 36th {Annual} {International} {Symposium}
  on {Computer} {Architecture}}, {ISCA} '09, pages 152--163, New York, NY, USA,
  2009. ACM.
\newblock ISBN 978-1-60558-526-0.
\newblock \doi{10.1145/1555754.1555775}.

\bibitem[Hong and Kim(2010)]{hong_integrated_2010}
S.~Hong and H.~Kim.
\newblock An integrated {GPU} power and performance model.
\newblock In \emph{{ACM} {SIGARCH} {Computer} {Architecture} {News}},
  volume~38, pages 280--289. ACM, 2010.

\bibitem[Joseph et~al.(2006)Joseph, Vaswani, and
  Thazhuthaveetil]{joseph_predictive_2006}
P.~J. Joseph, K.~Vaswani, and M.~J. Thazhuthaveetil.
\newblock A predictive performance model for superscalar processors.
\newblock In \emph{Proceedings of the 39th {Annual} {IEEE}/{ACM}
  {International} {Symposium} on {Microarchitecture}}, pages 161--170. IEEE
  Computer Society, 2006.

\bibitem[Klöckner(2014)]{kloeckner_loopy_2014}
A.~Klöckner.
\newblock Loo.{Py}: {Transformation}-based {Code} {Generation} for {GPUs} and
  {CPUs}.
\newblock In \emph{Proceedings of {ACM} {SIGPLAN} {International} {Workshop} on
  {Libraries}, {Languages}, and {Compilers} for {Array} {Programming}},
  {ARRAY}'14, pages 82:82--82:87, New York, NY, USA, 2014. ACM.
\newblock ISBN 978-1-4503-2937-8.
\newblock \doi{10.1145/2627373.2627387}.

\bibitem[Klöckner(2015)]{kloeckner_loopy_2015}
A.~Klöckner.
\newblock Loo.{Py}: {From} {Fortran} to {Performance} via {Transformation} and
  {Substitution} {Rules}.
\newblock In \emph{Proceedings of the 2Nd {ACM} {SIGPLAN} {International}
  {Workshop} on {Libraries}, {Languages}, and {Compilers} for {Array}
  {Programming}}, {ARRAY} 2015, pages 1--6, New York, NY, USA, 2015. ACM.
\newblock ISBN 978-1-4503-3584-3.
\newblock \doi{10.1145/2774959.2774969}.

\bibitem[Linderman et~al.(2008)Linderman, Collins, Wang, and
  Meng]{linderman_merge:_2008}
M.~D. Linderman, J.~D. Collins, H.~Wang, and T.~H. Meng.
\newblock Merge: {A} {Programming} {Model} for {Heterogeneous} {Multi}-core
  {Systems}.
\newblock In \emph{Proceedings of the 13th {International} {Conference} on
  {Architectural} {Support} for {Programming} {Languages} and {Operating}
  {Systems}}, {ASPLOS} {XIII}, pages 287--296, New York, NY, USA, 2008. ACM.
\newblock ISBN 978-1-59593-958-6.
\newblock \doi{10.1145/1346281.1346318}.

\bibitem[Meuer et~al.(2015)Meuer, Strohmaier, Dongarra, and
  Simon]{meuer2015top500}
H.~Meuer, E.~Strohmaier, J.~Dongarra, and H.~Simon.
\newblock Top500 supercomputing sites, November 2015.

\bibitem[Munshi et~al.(2011)Munshi, Gaster, Mattson, and
  Ginsburg]{munshi2011opencl}
A.~Munshi, B.~Gaster, T.~G. Mattson, and D.~Ginsburg.
\newblock \emph{OpenCL programming guide}.
\newblock Pearson Education, 2011.

\bibitem[{Nvidia Corporation}(2015)]{nvidia2015cuda}
{Nvidia Corporation}.
\newblock \emph{{CUDA C Programming Guide (Version 7.5)}}.
\newblock 2015.

\bibitem[Verdoolaege(2010)]{verdoolaege_isl_2010}
S.~Verdoolaege.
\newblock isl: {An} integer set library for the polyhedral model.
\newblock In \emph{Mathematical {Software}–{ICMS} 2010}, pages 299--302.
  Springer, 2010.

\bibitem[Verdoolaege et~al.(2007)Verdoolaege, Seghir, Beyls, Loechner, and
  Bruynooghe]{verdoolaege_counting_2007}
S.~Verdoolaege, R.~Seghir, K.~Beyls, V.~Loechner, and M.~Bruynooghe.
\newblock Counting integer points in parametric polytopes using {Barvinok}'s
  rational functions.
\newblock \emph{Algorithmica}, 48\penalty0 (1):\penalty0 37--66, 2007.

\bibitem[Williams et~al.(2009)Williams, Waterman, and
  Patterson]{williams_roofline_2009}
S.~Williams, A.~Waterman, and D.~Patterson.
\newblock Roofline: an insightful visual performance model for multicore
  architectures.
\newblock \emph{Communications of the ACM}, 52\penalty0 (4):\penalty0 65--76,
  2009.

\bibitem[Zhang and Owens(2011)]{zhang_quantitative_2011}
Y.~Zhang and J.~D. Owens.
\newblock A quantitative performance analysis model for {GPU} architectures.
\newblock In \emph{2011 {IEEE} 17th {International} {Symposium} on {High}
  {Performance} {Computer} {Architecture} ({HPCA})}, pages 382--393, Feb. 2011.
\newblock \doi{10.1109/HPCA.2011.5749745}.

\end{thebibliography}
\bibliographystyle{abbrvnat}

\end{document}